\documentclass[aps,prd,reprint,twocolumn,superscriptaddress,showpacs]{revtex4-1}
\usepackage{float}
\usepackage{booktabs}
\usepackage{siunitx}
\usepackage{graphicx}
\usepackage{mathrsfs}
\usepackage{bm}
\usepackage{amsmath}
\usepackage{dcolumn}
\usepackage{epstopdf}
\usepackage{dsfont}
\usepackage{amssymb}
\usepackage{tabularx}
\usepackage{array}
\usepackage{float}
\usepackage{color}
\usepackage{epstopdf}
\usepackage{mathrsfs}
\usepackage[colorlinks,linkcolor=blue,anchorcolor=blue,citecolor=blue,urlcolor=blue]{hyperref}
\usepackage{hyperref}
\usepackage{verbatim}

\begin{document}
\title{Mirror Chern insulators in two-dimensional altermagnetic Tc$_2$Cl$_2$O and Tc$_2$Br$_2$O}

\author{Rong Wang}
\affiliation{School of Physics, Northwest University, Xi'an 710127, China}

\author{Ruo-Yu Ning}
\affiliation{School of Physics, Northwest University, Xi'an 710127, China}

\author{Zhi-Hua Yan}
\affiliation{School of Physics, Northwest University, Xi'an 710127, China}

\author{Si Li}
\email{sili@nwu.edu.cn}
\affiliation{School of Physics, Northwest University, Xi'an 710127, China}
\affiliation{Shaanxi Key Laboratory for Theoretical Physics Frontiers, Xi'an 710127, China}
\affiliation{Peng Huanwu Center for Fundamental Theory, Xi'an 710127, China}
\affiliation{Fundamental Discipline Research Center for Quantum Science and Technology of Shaanxi Province, Xi'an 710127, China}
\begin{abstract}
	The interplay between altermagnetism and crystalline band topology provides
	an intriguing avenue for realizing unconventional topological phases with
	distinctive spin-dependent properties. Here, based on first-principles
	calculations and theoretical analysis, we identify monolayer
	$\mathrm{Tc}_2X_2\mathrm{O}$ ($X$ = Cl, Br) as a family of two-dimensional
	altermagnetic mirror Chern insulators. In the absence of spin--orbit
	coupling (SOC), both monolayers exhibit robust altermagnetism with
	mirror-spin coupling and host two symmetry-protected Weyl points in each
	spin channel near the Fermi level. The Weyl points in opposite spin channels carry distinct mirror-symmetry eigenvalues, $m_z=\pm i$.
	Upon inclusion of SOC, the Weyl points are gapped, and the two mirror
	sectors acquire opposite Chern numbers, ${\cal {C}}_{+}=1$ and ${\cal {C}}_{-}=-1$,
	resulting in a nonzero mirror Chern number ${\cal {C}}_m=1$. A low-energy
	$k\cdot p$ model captures the symmetry protection of the Weyl points and
	elucidates their SOC-induced mass gaps and topological character.
	Furthermore, the resulting mirror Chern insulating phases host helical
	edge states within the bulk band gap and exhibit a quantized spin Hall
	conductivity. Our work establishes a direct connection between
	altermagnetism and mirror Chern topology and provides a promising
	platform for exploring unconventional topological and spin-dependent
	phenomena in two-dimensional altermagnetic materials.
\end{abstract}

\maketitle
\section{Introduction}
Topological matter has attracted enormous interest and undergone rapid development in condensed matter physics over the past several decades~\cite{hasan2010colloquium,qi2011topological,bansil2016colloquium,chiu2016classification,tokura2019magnetic,armitage2018weyl,zhang2019catalogue,vergniory2019complete,tang2019comprehensive,xu2020high,lv2021experimental,vergniory2022all}. The field originated with the discovery of the quantum Hall effect and the Chern insulator (CI)~\cite{klitzing1980new,thouless1982quantized,haldane1988model}, both of which require broken time-reversal ($\mathcal{T}$) symmetry to generate a nonzero Berry curvature and the associated quantized Hall response. A subsequent major breakthrough was the discovery of symmetry-protected topological phases that do not rely on broken $\mathcal{T}$ symmetry, most notably the $\mathcal{T}$-invariant topological insulator~\cite{kane2005z,kane2005quantum} and the mirror Chern insulator (MCI)~\cite{teo2008surface,hsieh2012topological,tanaka2012experimental}. The $\mathcal{T}$-invariant topological insulator has since been realized in a wide range of materials, driving the rapid expansion of the field over the past two decades~\cite{hasan2010colloquium,qi2011topological,bansil2016colloquium,chiu2016classification}. Meanwhile, the MCI, protected instead by crystalline mirror symmetry, gave rise to the broader field of crystalline topological states and has since been extended to a rich family of symmetry-indicated topological phases, remaining an active area of research today~\cite{fu2011topological,zhang2019catalogue,vergniory2019complete,tang2019comprehensive}. Despite this progress, the interplay between crystalline topology and unconventional magnetic order remains comparatively unexplored, motivating the search for new material platforms in which these two ingredients coexist.

On the other hand, altermagnets (AMs) have attracted widespread attention as a newly identified fundamental magnetic phase, harmonizing zero net magnetization with non-relativistic, momentum-dependent spin splitting~\cite{smejkal2022conventional,smejkal2022emerging,wu2007fermi,hayami2019momentum,yuan2020giant,ma2021multifunctional,liu2022spin,bai2024altermagnetism,fender2025altermagnetism,song2025altermagnets}. Unlike conventional antiferromagnets, in which opposite-spin sublattices are typically related by spatial inversion or translation, the opposite-spin sublattices in AMs are instead connected by specific crystal rotations or mirror symmetries~\cite{smejkal2022conventional,smejkal2022emerging,liu2022spin,xiao2024spin,chen2024enumeration,jiang2024enumeration}. This unconventional symmetry relation enables spin degeneracy to be lifted even in the absence of spin–orbit coupling (SOC), giving rise to robust momentum-dependent spin splitting throughout the Brillouin zone (BZ), with $d$-, $g$-, or $i$-wave symmetry depending on the underlying crystal symmetry. Such a distinctive electronic structure underlies a broad range of unconventional phenomena, including the anomalous Hall effect~\cite{vsmejkal2020crystal,feng2022anomalous}, symmetry-driven spin-current generation~\cite{wu2007fermi,ma2021multifunctional,gonzalezhernandez2021efficient,bose2022tilted}, giant tunneling magnetoresistance~\cite{smejkal2022giant,shao2021spin}, spin Seebeck and Nernst effects~\cite{cui2023efficient}, and topological insulators~\cite{han2024cornertronics,wang2025pentagonal,wang2026two,wang2025real,zhang2026quantized}. Furthermore, AMs serve as a promising host for superconducting and multiferroic phenomena, including exotic Andreev reflections~\cite{papaj2023andreev,sun2023andreev}, finite-momentum Cooper pairing~\cite{zhang2024finite,hong2025unconventional,sim2025pair,chakraborty2024zero}, topological superconductivity~\cite{li2023majorana,ghorashi2024altermagnetic,li2024creation,zhu2023topological}, and (anti)ferroelectric orders~\cite{gu2025ferroelectric,yang2017topological,vsmejkal2024altermagnetic}. The symmetry-driven nature of altermagnetism naturally raises the question of whether altermagnetic order can be intertwined with crystalline band topology. In particular, mirror symmetry can both relate opposite-spin sublattices in altermagnets and protect MCIs, providing a natural bridge between these two concepts. Nevertheless, MCIs in altermagnetic systems remain largely unexplored. The realization of an altermagnetic MCI would therefore establish a direct link between unconventional magnetism and crystalline topology, offering a promising platform for exploring novel topological and spin-dependent phenomena.

Here, based on first-principles calculations and theoretical analysis, we demonstrate that monolayer Tc$_2$X$_2$O ($X$ = Cl, Br) realizes a novel class of altermagnetic MCIs. In the absence of SOC, monolayer Tc$_2$X$_2$O ($X$ = Cl, Br) exhibits altermagnetism with mirror-spin coupling and hosts two ideal Weyl points (WPs) in each spin channel near the Fermi level. Owing to the mirror-spin coupling, the WPs in opposite spin channels carry distinct $\mathcal{M}_z = \{C_{2z}^s \| M_z^l\}$ eigenvalues, $m_z=\pm i$. Upon inclusion of SOC, all WPs become gapped. The gapped WPs in the $m_z=+i$ and $m_z=-i$ mirror sectors contribute Chern numbers of $+1$ and $-1$, respectively, establishing monolayer Tc$_2$X$_2$O ($X$ = Cl, Br) as two-dimensional magnetic MCIs. Furthermore, we demonstrate that these MCIs host helical edge states within the bulk band gap, supporting dissipationless spin currents and giving rise to a quantized spin Hall conductivity plateau. A low-energy effective model is constructed to elucidate the nature of the WPs and the emergence of the mirror Chern insulating phases. Our findings establish a direct connection between altermagnetism and mirror Chern topology and provide a promising platform for exploring unconventional topological and spin-transport phenomena.

\section{First-principles Methods}
First-principles calculations were performed within the framework of density functional theory (DFT), as implemented in the Vienna \textit{ab initio} Simulation Package (VASP)~\cite{kresse1994ab,kresse1996efficient}. The exchange-correlation interactions were described using the Perdew--Burke--Ernzerhof (PBE) parametrization of the generalized gradient approximation (GGA)~\cite{perdew1996generalized}. The electronic wave functions were expanded in a plane-wave basis set with a kinetic energy cutoff of 600 eV, and Brillouin zone (BZ) integration was performed using a $\Gamma$-centered $12 \times 12 \times 1$ $k$-mesh. The energy and force convergence criteria were set to $10^{-5}$ eV and 0.01 eV/\AA, respectively. A vacuum layer of 15 \AA\ was introduced along the $z$ axis to eliminate spurious interactions between periodic images. Correlation effects among the Tc-$4d$ electrons were accounted for using the DFT+$U$ method~\cite{anisimov1991,dudarev1998}, with an effective $U$ value of 2 eV for the Tc atoms~\cite{moore2024high}. The Wilson loop spectra and edge states were computed via the iterative Green's function method~\cite{sancho1984quick,sancho1985highly}, as implemented in the WannierTools package~\cite{wu2018wanniertools}.

\section{RESULTS AND DISCUSSION}
	
\subsection{Crystal Structure and Magnetic Configurations}
\begin{figure}[htb]
	\includegraphics[width=8.6cm]{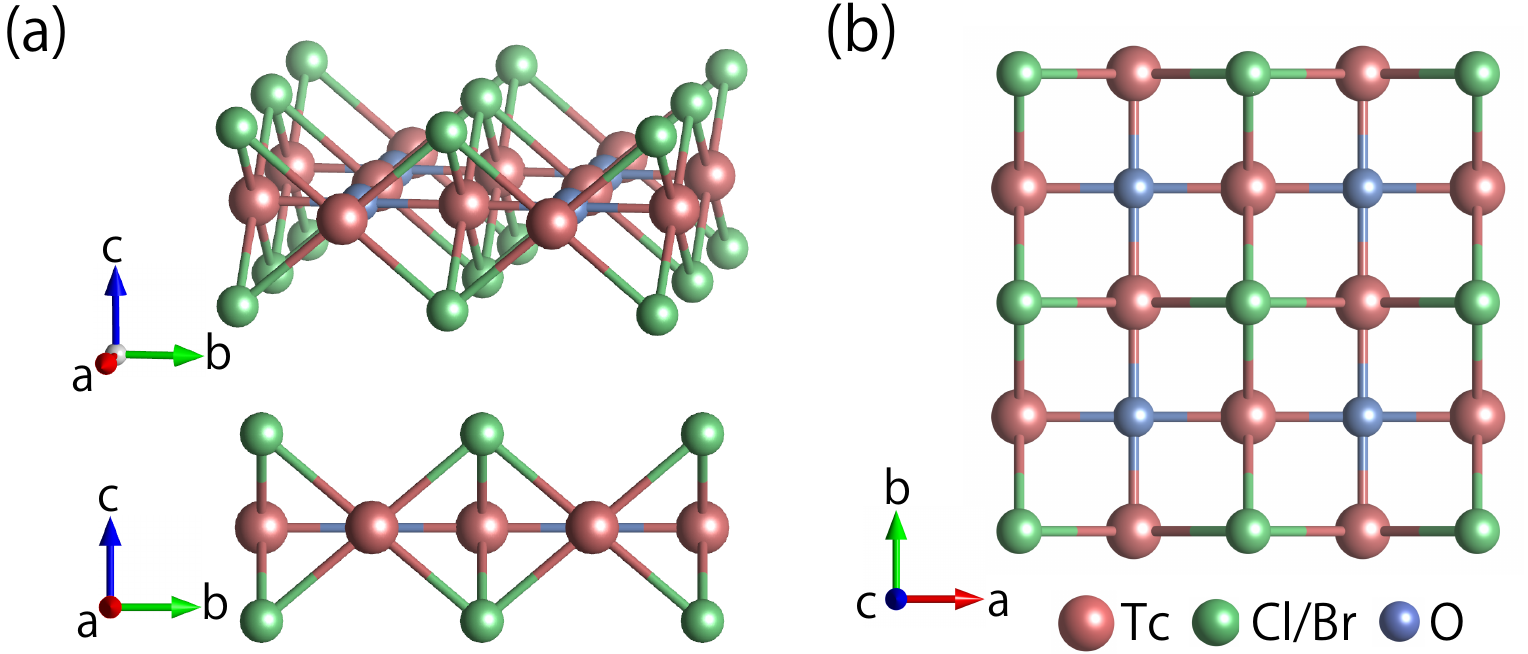}
	\caption{(a) Side and (b) top views of the crystal structure of monolayer Tc$_2X_2$O ($X$ = Cl, Br).}
	\label{fig1}
\end{figure}
The monolayer Tc$_2X_2$O ($X$=Cl, Br) adopts a tetragonal crystal structure consisting of three atomic layers stacked in the sequence $X$-(Tc-O)-$X$, as illustrated in Fig.~\ref{fig1}. The two Tc atoms and the O atom in the primitive cell are coplanar and are sandwiched between the two outer $X$ atomic layers. The crystal belongs to the tetragonal space group $P4/mmm$ (No.~123). The optimized lattice constants are $a=b=4.287$~\AA\ for Tc$_2$Cl$_2$O and $a=b=4.262$~\AA\ for Tc$_2$Br$_2$O, as summarized in Table~\ref{table1}.
\begin{figure*}[htb]
	\includegraphics[width=15cm]{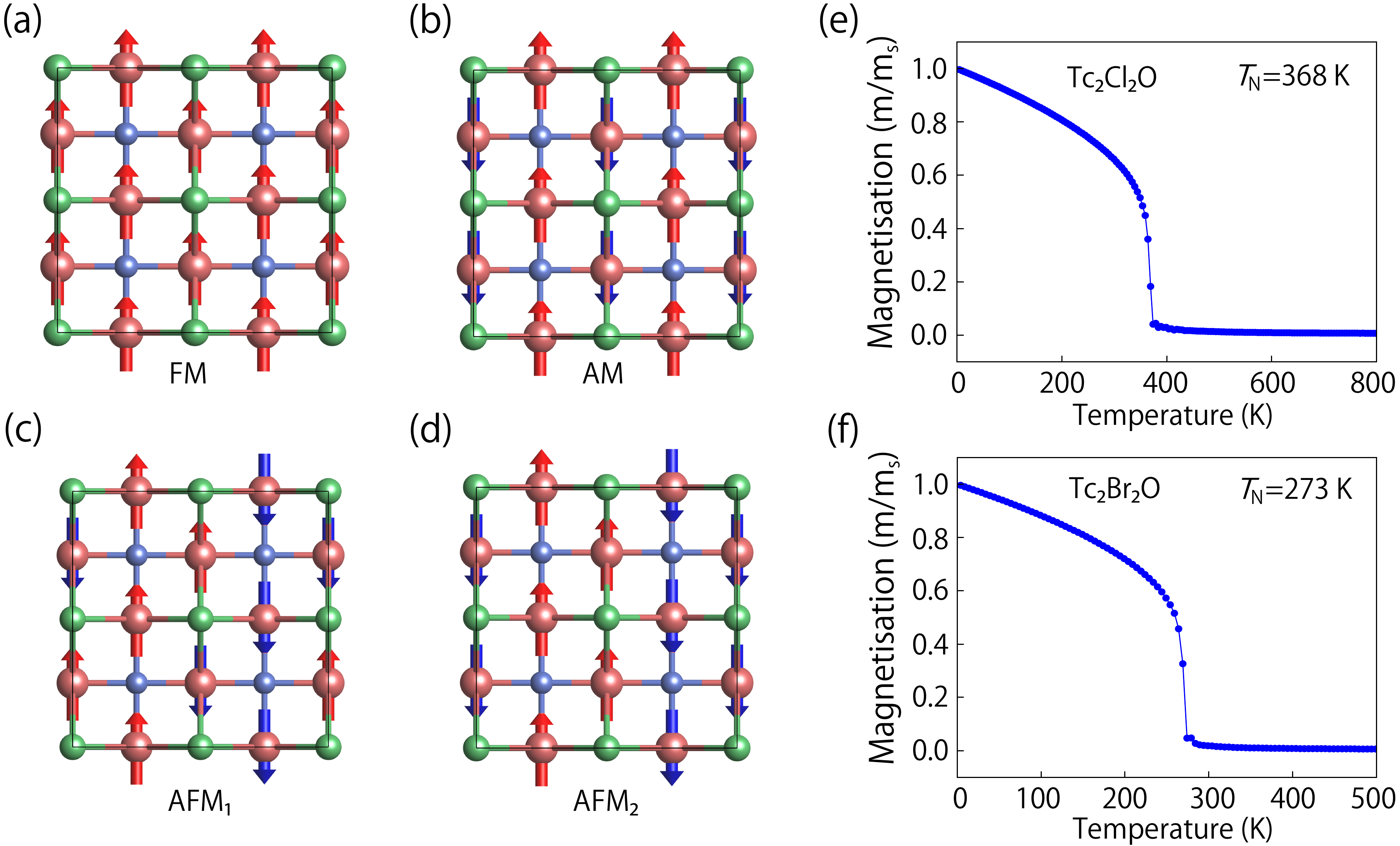}
	\caption{Panels (a)–(d) illustrate the FM, AM and two AFM configurations of the Tc$_2X_2$O ($X$ = Cl, Br) monolayer. Panels
		(e) and (f) show the normalized magnetic moments of monolayer Tc$_2X_2$O ($X$ = Cl, Br) as functions of temperature, obtained from Monte Carlo simulations. }
	\label{fig2} 
\end{figure*}

The partially filled Tc $4d$ orbitals give rise to intrinsic local magnetic moments in monolayer Tc$_2$X$_2$O. To determine the magnetic ground state, four representative magnetic configurations were considered [Fig.~\ref{fig2}(a)--(d)]: ferromagnetic (FM), altermagnetic (AM), and two antiferromagnetic configurations (AFM$_1$ and AFM$_2$). Total-energy calculations reveal that the AM configuration is the energetically most favorable state for both Tc$_2$Cl$_2$O and Tc$_2$Br$_2$O, with the corresponding relative energies summarized in Table~\ref{table1}. In the AM ground state, the magnetic moments are predominantly localized on the Tc atoms, with magnitudes of approximately 3.83~$\mu_B$ and 3.67~$\mu_B$ for Tc$_2$Cl$_2$O and Tc$_2$Br$_2$O, respectively. In the absence of SOC, the two spin sublattices are related by the $\{E^s||C_{4z}^l\}\mathcal{T}$ symmetry, confirming the altermagnetic nature of these monolayers.
To determine the magnetic anisotropy, we further performed SOC calculations, which reveal that both Tc$_2$Cl$_2$O and Tc$_2$Br$_2$O favor an out-of-plane N\'eel vector along the [001] direction. The corresponding magnetocrystalline anisotropy energies (MAEs), defined as $\Delta E_{\mathrm{MAE}} = E_{[001]} - E_{[100]}$, are summarized in Table~\ref{table1}.

\begin{table*}[htb]
	\centering
	\caption{\label{table1} Calculated structural, magnetic, and electronic properties of monolayer $\mathrm{Tc}_2\mathrm{Cl}_2\mathrm{O}$ and $\mathrm{Tc}_2\mathrm{Br}_2\mathrm{O}$, including the lattice constants ($a$), energy differences between the altermagnetic (AM) state and the ferromagnetic (FM) as well as two antiferromagnetic (AFM) states ($\Delta E_{\mathrm{AM-FM}}$, $\Delta E_{\mathrm{AM-AFM_1}}$, and $\Delta E_{\mathrm{AM-AFM_2}}$), nearest-neighbor exchange interaction ($J_1$), magnetocrystalline anisotropy energy ($\Delta E_{\mathrm{MAE}} = E_{[001]} - E_{[100]}$), and band gaps ($E_{\mathrm{g}}$) calculated with SOC.}
	\begin{tabular*}{\textwidth}{@{\extracolsep{\fill}} l c c c c c c c }
		\hline\hline
		Systems & $a$ (\AA)  & $\Delta E_{\mathrm{AM-FM}}$ (eV) & $\Delta E_{\mathrm{AM-AFM}_1}$ (eV) & $\Delta E_{\mathrm{AM-AFM}_2}$ (eV) & $J_1$ (eV)  & $\Delta E_{\mathrm{MAE}}$ (meV) & $E_{\mathrm{g}}$ (meV) \\
		\hline
		$\mathrm{Tc}_2\mathrm{Cl}_2\mathrm{O}$ & 4.287                                   & $-1.424$                     & $-1.214$                        & $-0.521$ & $-114.218$  & 3.546    & 0.021                    \\
		$\mathrm{Tc}_2\mathrm{Br}_2\mathrm{O}$ & 4.262                                   & $-1.017$                     & $-0.976$                        & $-0.323$ & $-81.572$  & 3.528   & 0.047                        \\
		\hline\hline
	\end{tabular*}
\end{table*}

To assess their magnetic stability, we further estimated the N\'eel temperature ($T_N$) by performing Monte Carlo (MC) simulations based on the effective spin Hamiltonian~\cite{evans2014atomistic}
\begin{equation}\label{Heisenberg}
	H=-\sum_{ i, j} J_{i j} \bm{S}_{i} \cdot \bm{S}_{j}-K\sum_{i}\left(S_{i}^{z}\right)^{2},
\end{equation}
where $\bm{S}_i$ denotes the normalized spin vector at Tc site $i$, $J_{ij}$ is the exchange coupling constant between sites $i$ and $j$, and $K$ is the magnetic anisotropy constant. All Hamiltonian parameters were extracted from first-principles calculations; the nearest-neighbor exchange coupling $J_1$ and the magnetic anisotropy constant $K$ are listed in Table~\ref{table1}. The N\'eel temperature was determined from the temperature dependence of the sublattice magnetization, as shown in Fig.~\ref{fig2}(e). The MC simulations yield $T_N$ values of approximately 368~K for Tc$_2$Cl$_2$O and 273~K for Tc$_2$Br$_2$O. These results demonstrate the robust thermal stability of the altermagnetic order and highlight the potential of these monolayers for high-temperature spintronic applications.

\subsection{Without SOC: AM Weyl Semimetals}
\begin{figure*}[htb]
	\includegraphics[width=17cm]{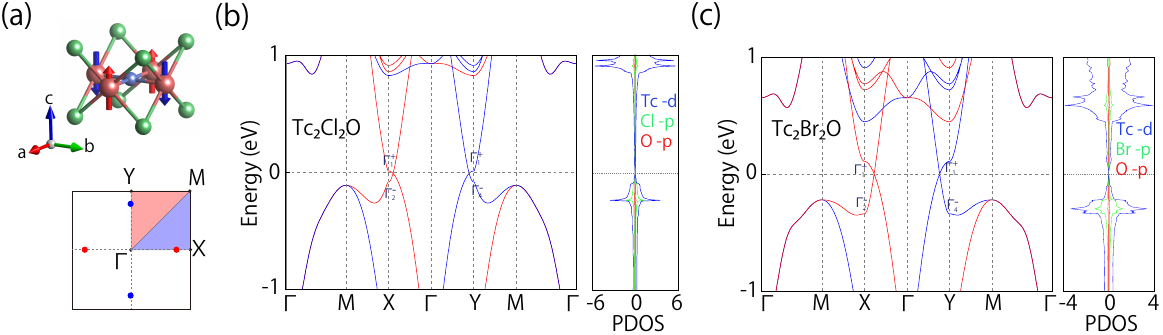}
	\caption{(a) Schematic illustrations of the altermagnetic configurations within the unit cells and the corresponding Brillouin zones of monolayer Tc$_2X_2$O ($X$ = Cl, Br). Red and blue dots indicate the positions of the Weyl points.. (b,c) Electronic band structures and projected density of states (PDOS) of monolayer Tc$_2$Cl$_2$O and Tc$_2$Br$_2$O without SOC, respectively. Red and blue curves represent the spin-up and spin-down states, respectively. Irreducible representations of the two bands near the Fermi level at the $X$ and $Y$ points are indicated.}
	\label{fig3}
\end{figure*}
Next, we investigate the electronic band structures of monolayer Tc$_2$X$_2$O ($X$ = Cl, Br) in their altermagnetic ground states.
In the absence of SOC, monolayer Tc$_2$X$_2$O ($X$=Cl, Br) belongs to spin space group No. 47.123.1.1. The system preserves the horizontal mirror symmetry ${\cal{M}}_z=\{M_z^s||M_z^l\}$, while the two spin-opposite sublattices are connected by the antiunitary operation ${\cal{C}}_{4z}=\{E^s||C_{4z}^l\}\mathcal{T}$ and by the mirror operations ${\cal{M}}_{\bar{1}10/110}=\{C_{2,\bar{1}10/110}^s||M_{\bar{1}10/110}^l\}$. Here, $C_{2z}^{s(l)}$, $P^{s(l)}$, and $M_z^{s(l)}$ denote the twofold rotation, inversion, and horizontal mirror operations in spin (lattice) space, respectively. Since inversion leaves the spin unchanged, the spin-space inversion operator is equivalent to the identity, i.e., $P^s \equiv E^s$.

The spin-resolved band structures of monolayer Tc$_2$Cl$_2$O and Tc$_2$Br$_2$O in the absence of SOC are shown in Figs.~\ref{fig3}(b) and \ref{fig3}(c), respectively. In both systems, the spin-up and spin-down bands exhibit pronounced momentum-dependent spin splitting, which displays a characteristic $d$-wave pattern and reverses sign between the $\Gamma$--X and $\Gamma$--Y directions, as required by the antiunitary symmetry ${\cal C}_{4z}$. Such momentum-dependent spin splitting in the absence of SOC is a hallmark of altermagnetism.
In addition, both monolayers exhibit semimetallic band structures featuring symmetry-protected Weyl points. Along the $\Gamma$--X path, a Weyl point arises from the crossing of two spin-up bands, whereas along the $\Gamma$--Y path, a Weyl point arises from the crossing of two spin-down bands. By virtue of the crystal symmetries, these crossings give rise to a total of two spin-up and two spin-down Weyl points within the Brillouin zone, with their positions shown in Fig.~\ref{fig3}(a).

We further demonstrate that the combined action of the mirror symmetry ${\cal M}_z$ and the antiunitary symmetry ${\cal C}_{4z}$ imposes a stringent constraint on the mirror characters of the Bloch states near the Weyl points.
Since ${\cal{M}}_z$ satisfies ${\cal{M}}_z^2=-1$, the Bloch states can be classified according to the mirror eigenvalues $m_z=\pm i$. We therefore choose the Bloch basis as the common eigenstates of ${\cal{M}}_z$, denoted by $|m_z,\mathbf{k}_{\uparrow(\downarrow)}\rangle$, where $\mathbf{k}_{\uparrow(\downarrow)}$ labels the crystal momentum of the spin-up (spin-down) state. Because ${\cal C}_{4z}$ commutes with ${\cal M}_z$,
\begin{eqnarray}
	[{\cal{M}}_z,{\cal{C}}_{4z}]=0,
\end{eqnarray}
acting with ${\cal{C}}_{4z}$ on a mirror eigenstate yields
\begin{eqnarray}
	{\cal M}_z{\cal C}_{4z}|m_z,\mathbf{k}_{\uparrow(\downarrow)}\rangle=
	m_z^{*} {\cal C}_{4z}|m_z,\mathbf{k}_{\uparrow(\downarrow)}\rangle,
\end{eqnarray}
where the complex conjugation of the mirror eigenvalue arises from the antiunitary nature of ${\cal C}_{4z}$. Since $m_z=\pm i$, it follows that $m_z^{*}=\mp i$, namely,
\begin{eqnarray}
	{\cal M}_z{\cal C}_{4z}|\pm i,\mathbf{k}_{\uparrow(\downarrow)}\rangle=
	\mp i\,{\cal C}_{4z}|\pm i,\mathbf{k}_{\uparrow(\downarrow)}\rangle.
\end{eqnarray}
Therefore, the spin-up and spin-down bands connected by ${\cal C}_{4z}$ necessarily possess opposite mirror eigenvalues, giving rise to a symmetry-enforced mirror-spin coupling. This symmetry analysis is fully corroborated by our DFT calculations, which confirm the opposite mirror characters of the two spin channels.

\subsection{With SOC: mirror Chern insulators}
\begin{figure*}[t]
	\includegraphics[width=15cm]{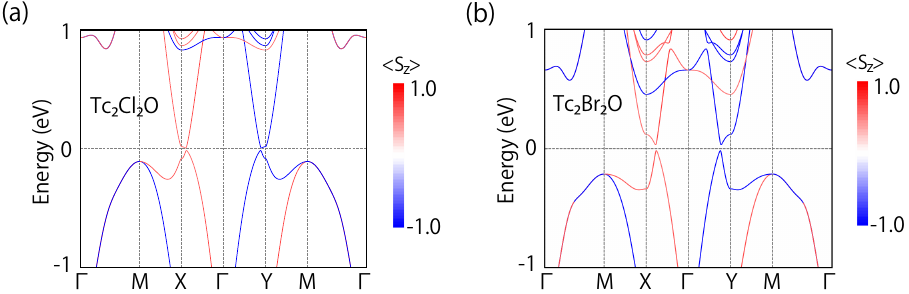}
	\caption{(a) and (b)  Band structures including SOC with the spin projection s$_z$ for monolayer Tc$_2$Cl$_2$O and Tc$_2$Br$_2$O monolayer, respectively. }
	\label{fig4} 
\end{figure*}
As established above, the inclusion of SOC stabilizes an out-of-plane N\'eel vector along the [001] direction. The corresponding magnetic symmetry is described by magnetic space group No. 123.4.1002, which preserves the mirror symmetry ${\cal M}_z$, the antiunitary symmetry $C_{4z}\mathcal{T}$, and the spatial inversion symmetry $\mathcal{P}$.
The band structures of monolayer Tc$_2$Cl$_2$O and Tc$_2$Br$_2$O with SOC are shown in Figs.~\ref{fig4}(a) and \ref{fig4}(b), respectively. It is evident that SOC opens small gaps at all the Weyl points, yielding band gaps of 0.021~eV for Tc$_2$Cl$_2$O and 0.047~eV for Tc$_2$Br$_2$O, respectively.

In the presence of SOC, the mirror symmetry ${\cal M}_z$ is preserved, and its eigenvalues remain $m_z=\pm i$. Consequently, the mirror-spin coupling persists, allowing the low-energy Bloch states to be classified according to their mirror eigenvalues, with the bands at $m_z=+i$ and $m_z=-i$ retaining opposite spin polarizations. Because ${\cal M}_z$ commutes with the full Hamiltonian, the Bloch Hamiltonian block-diagonalizes into two decoupled mirror subspaces, and a well-defined Chern number can be assigned to each subspace independently. This decomposition is precisely what enables the mirror Chern number, ${\cal {C}}_m$, to be defined as ${\cal {C}}_m=({{\cal C}_{+}-{\cal C}_{-}})/{2}$,
where ${\cal C}_{+}$ and ${\cal C}_{-}$ are the Chern numbers of the occupied states in the $m_z=+i$ and $m_z=-i$ mirror subspaces, respectively, calculated from
\begin{equation}
	{\cal C}_{\pm }=\frac{1}{2\pi}\int_{\mathrm{BZ}}\Omega(\mathbf{k})\,d^2k,
	\label{eq:CM}
\end{equation}
where the Berry curvature of each mirror subspace is given by
\begin{equation}
	\Omega(\mathbf{k})
	=
	\sum_{m\neq n}
	-2\,
	\mathrm{Im}
	\frac{
		\langle\psi_{n\mathbf{k}}|v_x|\psi_{m\mathbf{k}}\rangle
		\langle\psi_{m\mathbf{k}}|v_y|\psi_{n\mathbf{k}}\rangle
	}{
		(\varepsilon_{m\mathbf{k}}-\varepsilon_{n\mathbf{k}})^2
	},
	\label{eq:Berry}
\end{equation}
where $m$ and $n$ denote the band indices, $\varepsilon_{m\mathbf{k}}$ and $\varepsilon_{n\mathbf{k}}$ are the corresponding band energies, and $v_x$ and $v_y$ are the velocity operators. Physically, since each pair of spin-opposite Weyl points carries opposite mirror eigenvalues, SOC-induced gapping converts them into oppositely charged Berry-curvature sources within their respective mirror subspaces, thereby producing Chern numbers of equal magnitude but opposite sign.

\begin{figure*}[htp]
	\includegraphics[width=17.5cm]{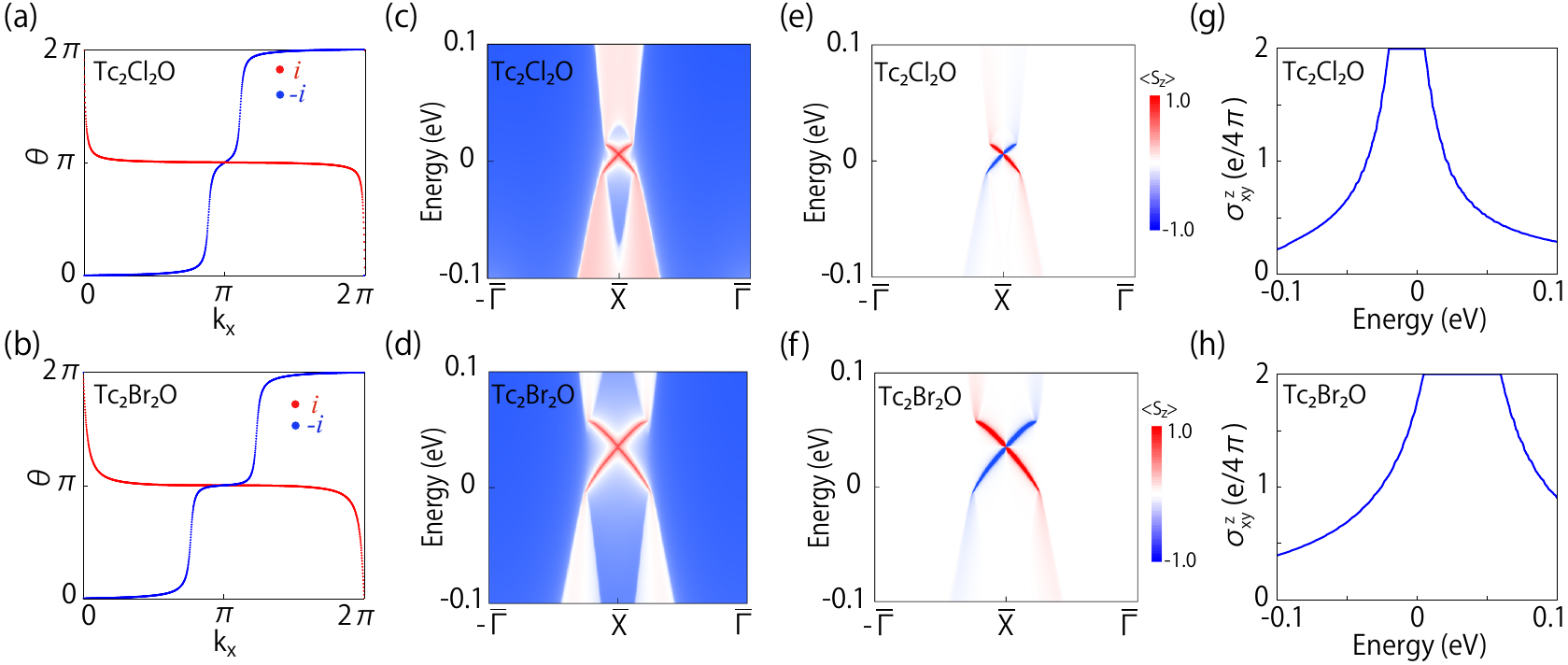}
	\caption{(a,b) Wilson loop evolutions for monolayer Tc$_2$Cl$_2$O and Tc$_2$Br$_2$O 
		with SOC, calculated in the two ${\cal M}_z$ ($m_z=\pm i$) subspaces, respectively. 
		(c,d) Edge-state spectra of monolayer Tc$_2$Cl$_2$O and Tc$_2$Br$_2$O along the 
		$(1\bar{1}0)$ direction, respectively. (e,f) Corresponding spin-resolved edge-state spectra with s$_z$ projection.
		(g,h) Energy dependence of the spin Hall conductivity $\sigma^{s}_{xy}$ for 
		monolayer Tc$_2$Cl$_2$O and Tc$_2$Br$_2$O, respectively, exhibiting a 
		quantized plateau within the bulk band gap.}
	\label{fig5} 
\end{figure*}

Our calculations yield ${\cal C}_{+}=1$ and ${\cal C}_{-}=-1$, giving an integer mirror Chern number of ${\cal {C}}_m=1$ for both monolayers. Notably, the total Chern number ${\cal C}={\cal C}_{+}+{\cal C}_{-}=0$, confirming that the overall time-reversal-broken state remains a Chern-trivial insulator and that its nontrivial topology is protected exclusively by the crystalline mirror symmetry ${\cal M}_z$, rather than by a net Berry curvature integrated over the entire Brillouin zone. To further confirm the nontrivial topology, we calculated the Wilson loops for the two mirror subspaces separately for monolayer Tc$_2$Cl$_2$O and Tc$_2$Br$_2$O, as shown in Figs.~\ref{fig5}(a) and \ref{fig5}(b), respectively. In both subspaces, the Wilson loop spectra exhibit a clear winding as a function of $k_x$, unambiguously signaling a nonzero Chern number and providing independent confirmation consistent with the direct Berry-curvature integration. These results establish that monolayer Tc$_2$Cl$_2$O and Tc$_2$Br$_2$O are two-dimensional MCIs protected by the mirror symmetry ${\cal M}_z$ in the presence of SOC, representing a rare realization of MCI physics within an altermagnetic host.

A hallmark of a MCI is the existence of gapless edge states protected by mirror symmetry, in which the counterpropagating edge modes carry opposite mirror eigenvalues. To demonstrate this feature, we calculated the edge spectrum of a nanoribbon oriented along the $(1\bar{1}0)$ direction for monolayer Tc$_2$X$_2$O ($X=\mathrm{Cl},\mathrm{Br}$) with SOC. As shown in Figs.~\ref{fig5}(c) and \ref{fig5}(d), each edge hosts a pair of counterpropagating edge states that traverse the bulk gap and cross linearly at the $\mathrm{\bar{X}}$ point, in accordance with the mirror Chern number ${\cal C_M}=1$ obtained above. Remarkably, the two edge modes are almost completely polarized into opposite spin channels, as evidenced by the spin-resolved edge spectrum [see Figs.~\ref{fig5}(e) and \ref{fig5}(f)].
Although SOC is present, the hybridization between the two counterpropagating edge states remains negligible even as they approach each other in momentum space near $\mathrm{\bar{X}}$. This robustness originates from the mirror-spin coupling enforced by the mirror symmetry ${\cal M}_z$, which assigns opposite mirror eigenvalues to the two edge modes. Consequently, the crossing at $\mathrm{\bar{X}}$ is not merely accidental but is symmetry-enforced, guaranteeing a gapless, dissipationless conduction channel as long as ${\cal M}_z$ is preserved. This mirror-protected, spin-polarized edge crossing constitutes direct real-space evidence for the nontrivial mirror Chern number and, together with the Wilson-loop winding discussed above, firmly establishes monolayer Tc$_2$X$_2$O ($X=\mathrm{Cl},\mathrm{Br}$) as an altermagnetic MCI hosting helical, spin-polarized edge transport.

Finally, we investigate the intrinsic spin Hall conductivity, $\sigma_{xy}^{s}$, of the Tc$_2$X$_2$O ($X=\mathrm{Cl},\mathrm{Br}$) monolayers. Within linear-response theory, $\sigma_{xy}^{s}$ is evaluated using the Kubo formula~\cite{yao2005sign,sinova2015spin},
\begin{equation}
	\sigma_{xy}^{s}
	=
	e\hbar
	\int
	\frac{d^{2}k}{(2\pi)^2}
	\Omega^{s}(\mathbf{k}),
\end{equation}
where the spin Berry curvature is given by
\begin{equation}
	\Omega_n^{s}(\mathbf{k})
	=
	-2\,
	\mathrm{Im}
	\sum_{m\neq n}
	\frac{
		\langle\psi_{m\mathbf{k}}|J_x^{s}|\psi_{n\mathbf{k}}\rangle
		\langle\psi_{n\mathbf{k}}|v_y|\psi_{m\mathbf{k}}\rangle
	}{
		(\varepsilon_{n\mathbf{k}}-\varepsilon_{m\mathbf{k}})^2
	}.
	\label{eq:SHC}
\end{equation}
Here, $J_x^{s}=\frac{\hbar}{4}\{\sigma_z,v_x\}$ is the spin-current operator describing a spin current flowing along the $x$ direction with spin polarization along the $z$ axis. The calculated spin Hall conductivities, as functions of the Fermi energy, are presented in Figs.~\ref{fig5}(g) and \ref{fig5}(h). Within the SOC-induced bulk gap, $\sigma_{xy}^{s}$ exhibits an almost quantized plateau with a value close to $e/(2\pi)$. This nearly quantized spin Hall response stems from the nontrivial mirror topology characterized by the mirror Chern number ${\cal {C}}_m=1$. Owing to the mirror-spin coupling, the two mirror subspaces ($m_z=\pm i$) carry opposite spin polarizations and opposite Chern numbers, together giving rise to a robust spin Hall current protected by the mirror symmetry.

\subsection{Low-energy effective model}	

To further characterize the Weyl semimetal phase and to elucidate the effect of SOC, we construct a low-energy effective ($k\cdot p$) model based on symmetry considerations.

\begin{figure*}[htp]
	\includegraphics[width=13cm]{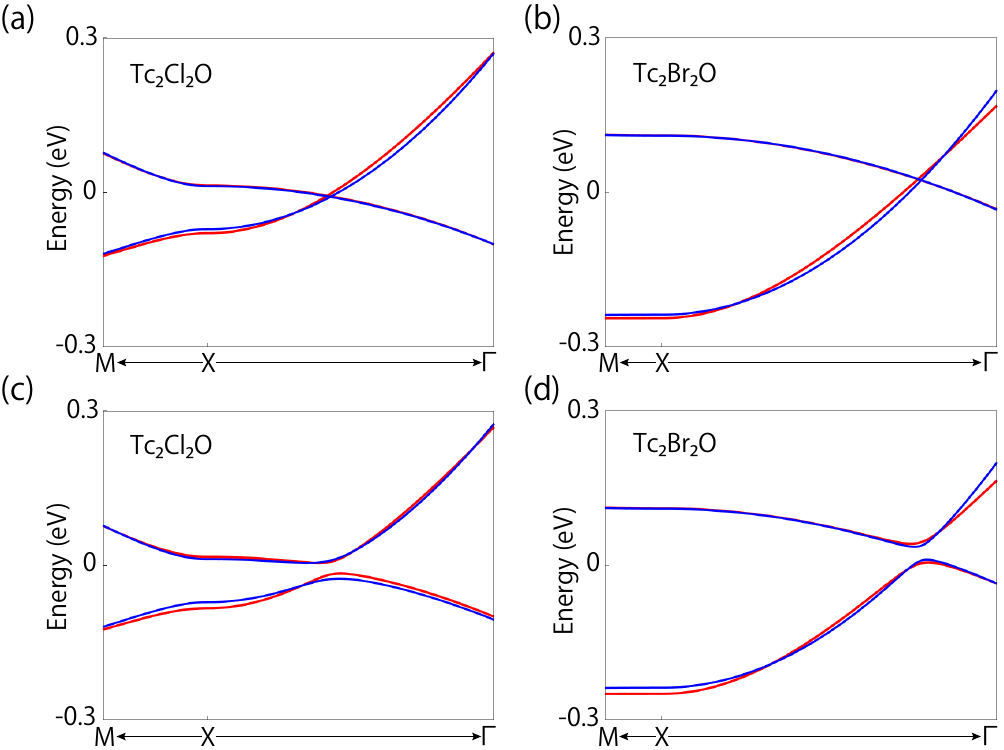}
	\caption{Dispersions around the Weyl points fitted by the effective model. Panels (a) and (b) show the band structures without SOC, while panels (c) and (d) show those with SOC. Red and blue curves represent the DFT results and the fitting results obtained from the $k\cdot p$ models in Eqs.~(\ref{kpnosoc}) and (\ref{kpsoc}), respectively.}
	\label{fig6} 
\end{figure*}

We first consider the case without SOC. For the spin-up channel, the two low-energy states at the $X$ point transform as the $\Gamma_2^{-}$ and $\Gamma_1^{+}$ irreducible representations of the little group $D_{2h}$ at $X$. Using these two states as the basis, we construct a symmetry-constrained two-band $k\cdot p$ Hamiltonian. To capture the Weyl point in the vicinity of $X$, we expand the Hamiltonian to leading order in each wave-vector component $k_i$, yielding
\begin{equation}\label{kpnosoc}
	\mathcal{H}(\mathbf{k}) =\left(t_x k_x^2 + t_y k_y^2 \right) \sigma_0+ \left( m + r_x k_x^2 + r_y k_y^2 \right) \sigma_z + v_y k_y \sigma_x,
\end{equation}
where the Pauli matrices $\sigma_i$ act within the space spanned by the two basis states at $X$, and $\sigma_0$ is the $2\times2$ identity matrix. The model parameters, obtained by fitting to the first-principles band structure, are $t_x=3.413$~eV\,\AA$^2$, $t_y=4.484$~eV\,\AA$^2$, $m=-0.418$~eV, $r_x=6.786$~eV\,\AA$^2$, $r_y=-0.001$~eV\,\AA$^2$, and $v_y=-2.019$~eV\,\AA\ for Tc$_2$Cl$_2$O, and $t_x=3.775$~eV\,\AA$^2$, $t_y=3.917$~eV\,\AA$^2$, $m=-0.220$~eV, $r_x=7.493$~eV\,\AA$^2$, $r_y=-0.152$~eV\,\AA$^2$, and $v_y=-1.888$~eV\,\AA\ for Tc$_2$Br$_2$O. The resulting fits to the DFT band structures are shown in Figs.~\ref{fig6}(a) and (b), confirming that this minimal two-band model faithfully captures the low-energy physics near $X$.
This model predicts two Weyl points along the $\Gamma$--$X$ path, which are protected by the \(\{E^s\|C_{2x}^l\}\) symmetry. Explicitly, the WPs are located at $\boldsymbol{k}_w = ( \pm\sqrt{|m/r_x|}, 0)$, as shown in Fig.~\ref{fig3}(a). Expanding Eq.~(\ref{kpnosoc}) around $\boldsymbol{k}_w$ and retaining terms linear in the momentum deviation $\boldsymbol{q}=(q_x,q_y)$, the effective Hamiltonian describing each WP reduces to
\begin{equation}\label{kpnosocWP}
	H_{\pm}^{\uparrow} = \pm 2\sqrt{|m/r_x|}\, r_x q_x \sigma_z + v_y q_y \sigma_x,
\end{equation}
which describes a Weyl point with linear dispersion along both $q_x$ and $q_y$. By the antiunitary symmetry $\{E^s\|C_{4z}^l\}\mathcal{T}$, the spin-up WPs $W_{\pm}^{\uparrow}$ are mapped onto a corresponding pair of spin-down WPs $W_{\pm}^{\downarrow}$, which are in turn protected by the symmetry $\{E^s\|C_{2y}^l\}$. Together, these four WPs constitute the minimal set of symmetry-required band-touching points in the spin-resolved band structure, consistent with the first-principles results discussed above.

In the presence of SOC, the two spin channels are no longer independently conserved, and an additional symmetry-allowed term couples them. The Hamiltonian in Eq.~(\ref{kpnosoc}) is accordingly modified to
\begin{align}\label{kpsoc}
	\mathcal{H}(\mathbf{k}) &= \left(t_x k_x^2 + t_y k_y^2\right) \sigma_0+ \left( m + r_x k_x^2 + r_y k_y^2 \right) \sigma_z \nonumber \\
	&\quad + v_y k_y \sigma_x + v_x k_x \sigma_y,
\end{align}
where the new term $v_xk_x\sigma_y$ is permitted by symmetry once SOC is included. Fitting to the first-principles band structures yields $v_x=-0.229$~eV\,\AA\ for Tc$_2$Cl$_2$O and $v_x=-0.107$~eV\,\AA\ for Tc$_2$Br$_2$O. The fitted band structures with SOC are shown in Figs.~\ref{fig6}(c) and (d). Expanding this Hamiltonian around $\boldsymbol{k}_w$, Eq.~(\ref{kpnosocWP}) generalizes to
\begin{equation}\label{kpsocWP}
	H_{\pm}^{\uparrow} = \pm 2\sqrt{|m/r_x|}\, r_x q_x \sigma_z + v_y q_y \sigma_x + v_x q_x \sigma_y \pm \sqrt{|m/r_x|}\,v_x\sigma_y,
\end{equation}
where the momentum-independent term $\pm\sqrt{|m/r_x|}\,v_x\sigma_y$ acts as a SOC-induced mass term that gaps out the Weyl point. Diagonalizing this two-band model shows that the resulting Chern number of each gapped WP is given by
${\cal {C}}_\pm=\frac{1}{2}\,\mathrm{Sign}\!\left(r_xv_xv_y\right)$,
which is identical for both $H_{+}^{\uparrow}$ and $H_{-}^{\uparrow}$, since the sign dependence on $\boldsymbol{k}_w$ cancels between the linear $\sigma_z$ coefficient and the mass term. Consequently, the two spin-up WPs contribute equally to the total Chern number of the spin-up channel, and SOC drives the spin-up ($m_z=i$) sector into a Chern insulator with ${\cal {C}}_+=1$. Because the two spin channels are related by the antiunitary symmetry $C_{4z}\mathcal{T}$, which reverses the sign of the Berry curvature, the spin-down ($m_z=-i$) channel must likewise become a Chern insulator, but with the opposite Chern number ${\cal {C}}_-=-1$. This symmetry-based prediction is independently corroborated by the Wilson-loop calculations performed separately for the two mirror subsystems, as shown in Figs.~\ref{fig5}(a) and \ref{fig5}(b). Taken together, the effective model establishes that monolayer Tc$_2$Cl$_2$O and Tc$_2$Br$_2$O, in the presence of SOC, realize a two-dimensional MCI with mirror Chern number ${\cal {C}}_m=({\cal {C}}_+-{\cal {C}}_-)/2=1$.

\section{Conclusion}
In summary, using first-principles calculations and symmetry-based analysis, we have identified monolayer $\mathrm{Tc}_2X_2\mathrm{O}$ ($X$ = Cl, Br) as a novel family of two-dimensional altermagnetic MCIs. In the absence of SOC, both systems exhibit altermagnetic order with momentum-dependent spin splitting and mirror-spin coupling, hosting two pairs of ideal Weyl points near the Fermi level whose opposite spin channels carry distinct mirror eigenvalues. Upon inclusion of SOC, all Weyl points become gapped, yielding a nonzero mirror Chern number and thereby establishing a genuinely magnetic mirror Chern insulating phase. This nontrivial topology is further corroborated by helical edge states that traverse the bulk gap, support dissipationless spin transport, and give rise to a quantized spin Hall conductivity plateau. Our results demonstrate that altermagnetic order and mirror-protected band topology can coexist, establishing $\mathrm{Tc}_2X_2\mathrm{O}$ monolayers as a promising platform for exploring the interplay between unconventional magnetism and topology, with potential applications in spin-dependent transport and spintronics.

\bigskip
\begin{acknowledgements}
	This work was supported by the Key Program of the Natural Science Basic Research Plan of Shaanxi Province (Grant No. 2025JC-QYCX-007) and the Youth Project (Category B) of the Natural Science Basic Research Plan of Shaanxi Province (Grant No. 2026JC-YXQN-026).
\end{acknowledgements}

%

\end{document}